\documentclass{article}

\usepackage{arxiv}
\usepackage[utf8]{inputenc} % allow utf-8 input
\usepackage[T1]{fontenc}    % use 8-bit T1 fonts       % hyperlinks
\usepackage{url}            % simple URL typesetting
\usepackage{booktabs}       % professional-quality tables
\usepackage{amsfonts}       % blackboard math symbols
\usepackage{microtype}      % microtypography
\usepackage{graphicx}
\usepackage{doi}
\usepackage{siunitx}
\usepackage{amsmath}
\usepackage{amssymb}
\usepackage{derivative}
\usepackage{lineno}
\usepackage{placeins}

\title{Do Inner Greenland's Melt Rate Dynamics Approach Coastal Ones?}

\author{\href{https://orcid.org/0000-0003-0529-7926}{\includegraphics[scale=0.06]{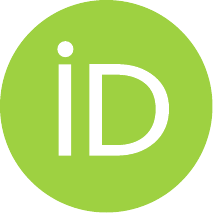}\hspace{1mm}Martin ~Heßler}\thanks{Center for Nonlinear Science, University of Münster, 48149 Münster, North Rhine-Westphalia, Germany} \\
	Institute of Theoretical Physics\\
	University of Münster\\
	48149 Münster, North Rhine-Westphalia, Germany\\
	\texttt{m\_{}hess23@uni-muenster.de} \\
    \And
	\href{https://orcid.org/0000-0003-0986-0878}{\includegraphics[scale=0.06]{orcid.pdf}\hspace{1mm}Oliver ~Kamps} \\
	Center for Nonlinear Science\\
	University of Münster\\
	48149 Münster, North Rhine-Westphalia, Germany\\
	\texttt{okamp@uni-muenster.de} \\
}

\renewcommand{\headeright}{\emph{arXiv} Preprint}
\renewcommand{\undertitle}{\emph{arXiv} Preprint}
\renewcommand{\shorttitle}{Greenland Melt Rate Dynamics}

\hypersetup{
pdftitle={Do Inner Greenland's Melt Rate Dynamics Approach Coastal Ones?},
pdfsubject={nlin.AO, physics.ao-ph, physics.comp-ph, physics.data-an},
pdfauthor={Martin Heßler, Oliver Kamps},
pdfkeywords={Melt Rates, Greenland Ice Sheet, Bayesian Langevin Estimation, Tipping, Resilience Analysis, Complex Systems},
}

\usepackage[sort&compress]{natbib}
\setcitestyle{numbers,square}
\usepackage{hyperref}

\makeatletter
\providecommand{\NAT@force@numbers}{} % Verhindern des Fehlers
\makeatother

\begin{document}
\maketitle

\begin{abstract}
The Greenland Ice Sheet may be nearing a tipping point, transitioning to permanent melting. This article analyses two melt rate time series using the Bayesian Langevin estimation (BLE), providing further evidence for destabilizing melt dynamics, along with new insights from the method's nonlinear parameterisation. Comparing the results for Western Central Greenland and the Nuussuaq peninsula suggests that inland melt dynamics may approach less stable coastal dynamics on slow and fast scales. Both datasets show a significant increase in fast-scale amplitudes since the 1970s, possibly driven by large-scale atmospheric fluctuations. Additionally, the BLE's nonlinear drift proves crucial for obtaining these results, as a linear Ornstein-Uhlenbeck process fails to capture these dynamics due to the oversimplification of a strictly positive autocorrelation function.
\end{abstract}

\keywords{Melt Rates \and Greenland Ice Sheet \and Bayesian Langevin Estimation \and Tipping \and Resilience Analysis \and Complex Systems}

\section{Introduction}
Climate change is one of the most critical issues of the 21st century and a major focus in the global scientific community. Without a doubt, the Earth's climate is a complex system, for which nonlinear modelling and empirical time series analysis have become essential tools to deepen our understanding.\\
Building on other pioneering works in the field, Svante Arrhenius~\cite{a:Arrhenius1896} provided the first explanation of the greenhouse effect in 1896. Since then, many phenomena and dependencies of the Earth’s climate have been profoundly studied, often focusing on climate subsystems~\cite{a:Lenton2008}, such as the Greenland Ice Sheet (GrIS)~\cite{a:Boers2021}.\\
The GrIS is believed to be nearing a tipping point or may potentially already be in a transient phase beyond it. In 2021, Boers and Rypdal~\cite{a:Boers2021} analysed melt rate time series extracted from stacked ice cores from Central Western Greenland (CWG) and an ice core from the Nuussuaq (NU) peninsula. Their results imply that the GrIS is threatened by a nearby tipping point at which its current state is no longer maintainable. Furthermore, the primary driver of the changing dynamics is identified as the melt-elevation feedback, i.e., the increase in melt rates with decreasing ice sheet height.\\
They utilized the non-parametric statistical lag-1 autocorrelation (AR1) $\rho_{1}$ and standard deviation (STD) $\hat{\sigma}_{\rm d}$ and found significant positive trends in both time series. In Section~\ref{subsec: melt rate analysis}, we complement these analyses by the parametric Bayesian Langevin estimation (BLE)~\cite{a:Hessler2021, a:Hessler2022}, which adds further evidence for decreasing resilience, along with new insights. The results suggest that the inland melt rate dynamics from CWG started to destabilize around 1914, approaoching the less stable niveau of coastal areas around 1975. Additionally, a recent increase of fast-scale contributions in the late 1990s in both CWG and the NU peninsula may be driven by a common fast-scale driver such as large-scale fluctuations in the atmospheric circulation. In particular, the results and implications of the BLE analysis for both datasets are not reproducible using the transformation from the AR1 and STD into the quantitative metrics of the Ornstein-Uhlenbeck estimation (OUE) (cf. Section~\ref{subsec: greenland BLE OUE}). This limitation arises due to the multidecadal variability~\cite{a:Ghil1991} present in the GrIS melt rates and the overall notably less accurate OUE. These findings are illustrated by a comparison between the BLE and the OUE in Section~\ref{subsec: greenland BLE OUE}. For a streamlined presentation, both the OUE and BLE are described beforehand in Section~\ref{sec: methods}. 

\section{Methods}
\label{sec: methods}
For the melt rate dynamics, we heuristically assume an autonomous Langevin equation
\begin{align}
    \dot{x} = h(x)+g(x)\Gamma (t),
\end{align}
where $h(x)$ and $g(x)$ denote the deterministic drift and stochastic diffusion, respectively, and $\Gamma (t)=\odv{W}{t}$ represents increments of a Wiener process~\cite{b:kloeden92}. A change of the sign in the slope 
\begin{align}
\zeta = \left. \frac{\text{d}h(x)}{\text{d}x}\right\vert_{x = x^*} 
\end{align}
of the nonlinear drift at the fixed point $x^*$ indicates destabilization of the fixed point through control parameter change, i.e., a bifurcation. The fixed point $x^*$ is estimated as mean in each window. The most straightforward approach is the linear approximation of the drift resulting in the OUE introduced in Section~\ref{subsec: OUE method}. However, most problems in nature are inherently nonlinear, and it is worthwhile to find numerical methods that reliably infer these nonlinearities from time series. The BLE~\cite{a:Hessler2021} provides a promising pathway for this purpose. Therefore, a short remainder of the method is given in Section~\ref{subsec: BLE method}. Since the melt rate dynamics are supposed to change over time, we apply the following methods to rolling data windows, which are assumed to be nearly stationary. 
\subsection{A First Approximation: The Ornstein-Uhlenbeck Estimation}
\label{subsec: OUE method}
When a complex system approaches a bifurcation, this can be accompanied by a decreasing restoring rate and conversely by an increasing relaxation time. This critical slowing down (CSD) is the starting point of non-parametric leading indicator argumentations: An equidistantly sampled time series should exhibit positive trends in AR1 $\hat{\rho}_{1}$ and STD $\hat{\sigma}_{\rm d}$ over time because perturbations persist longer when the restoring rate decreases due to CSD. Here, we focus on a common variant of parametric reasoning which involves the statistical measures AR1 $\hat{\rho}_{1}$ and STD $\hat{\sigma}_{\rm d}$. The data-generating process is approximated as an OU process
\begin{align*}
\dot{x} = \zeta_{\rm OU}\cdot x+\sigma_{\rm OU}\Gamma (t),
\end{align*}
i.e., a Langevin equation with a linear drift $h(x)=\zeta_{\rm OU}\cdot x$ and constant diffusion $g(x)=\text{const.}\equiv\sigma_{\rm OU}$. The AR1~$\rho_{1}$ and STD~$\hat{\sigma}_{\rm d}$ of an OU process realisation is given by 
\begin{align}
\label{eq:OU statistics}
\begin{split}
\rho_{1} &= \exp(\zeta_{\rm OU}\Delta t)\\
\sigma_{\rm d}^2 &= -\frac{\sigma^2_{\rm OU}}{2\zeta_{\rm OU}}
\end{split} 
\end{align}
with the discrete time step $\Delta t$~\cite{a:Morr2023}. Note that the non-parametric leading indicators implicitly build on such  dependencies $\rho_{1}(\zeta_{\rm OU})$ and $\sigma_{\rm d}^2(\zeta_{\rm OU})$, but do not parameterise a specific model. For pure B-tipping, all parameters of Equations~\ref{eq:OU statistics}, apart from the restoring force, are assumed to be constant. Let us exemplarily illustrate this using the OU model with restoring rate $\zeta_{\rm OU}$ (which is not necessarily specified in the non-parametric ansatz). Emerging CSD can be described as follows:
\begin{align*}
\lim\limits_{\zeta_{\rm OU}\to 0}\rho_{1}(\zeta_{\rm OU})&=1, \qquad \text{indicating a positive trend }\rho_{1}\nearrow\\
\lim\limits_{\zeta_{\rm OU}\to 0}\sigma_{\rm d}^2(\zeta_{\rm OU})&\rightarrow\infty ,\qquad \text{indicating a positive trend }\hat{\sigma}_{\rm d}^2\nearrow.
\end{align*}
However, in the parametric ansatz, we explicitly define the OU model to approximate the data-generating process. This means, once the AR1 $\hat{\rho}_{1}$ and STD $\hat{\sigma}_{\rm d}$ of a time series are estimated in rolling windows, they can be transformed through Relations~\ref{eq:OU statistics} into the restoring rate estimates $\hat{\zeta}_{\rm OU}$ and noise level estimates $\hat{\sigma}_{\rm OU}$ of an OU process by
\begin{align}
\label{eq:OU indicator}
\begin{split}
\hat{\zeta}_{\rm OU} &= \frac{\log(\hat{\rho}_{1})}{\Delta t}\\
\hat{\sigma}_{\rm OU} &= \sqrt{-2\hat{\sigma}_{\rm d}^2\hat{\zeta}_{\rm OU}} = \sqrt{\frac{-2\hat{\sigma}_{\rm d}^2\log(\hat{\rho}_{1})}{\Delta t}}.
\end{split}
\end{align}
Since the autocorrelation function $\rho_{\rm OU}(\tau)$ of an OU process satisfies $\rho(\tau)>0\;\forall\;\tau$ for all time lags $\tau$, the term $\log(\hat{\rho}_{1})$---which enters in both quantities of Equations~\ref{eq:OU indicator}---is only defined as long as the OU approximation is sufficiently met. Generally, for time series of arbitrary processes it holds $\rho(\tau)\in [-1,1]$.

\subsection{A Nonlinear Approximation: The Bayesian Langevin Estimation}
\label{subsec: BLE method}
In this section, the BLE, introduced in the original articles~\cite{a:Hessler2021, a:Hessler2022}, is briefly sketched. User-friendly implementations of the methods can be found in the Python package \textit{antiCPy}~\cite{url:DocsHessler2021, url:GitHessler2021}.\\
In the BLE, the drift $h(x)$ is expanded into a third-order Taylor series which is sufficient to describe the normal forms of simple bifurcation scenarios~\cite{b:strogatz}. Furthermore, in cases of strong noise the first approximation of small disturbances, i.e., $\mathcal{O}((x-x^*)^2)\approx 0$, breaks down and the BLE is more reliable due to the higher-order Taylor expansion~\cite{a:Hessler2022}. This results in
\begin{align}
\begin{split}
h(x(t),t) = \alpha_0(t) + \alpha_1(t)\cdot (x - x^*) + \alpha_2(t)\cdot (x - x^*)^2
 + \alpha_3(t)\cdot (x - x^*)^3 + \mathcal{O}((x-x^*)^4)  \label{eq:taylor} 
\end{split}
\end{align}
so that the information on the linear stability is incorporated in $\alpha_1$. For practical reasons in the numerical approach, Equation~\ref{eq:taylor} is used in the form
\begin{align}
\begin{split}
    h'_{\rm MC}(x(t),t) = \theta_0 (t; x^*) + \theta_1 (t; x^*) \cdot x + \theta_2 (t; x^*) \cdot x^2 + \theta_3 (t; x^*) \cdot x^3,
    \end{split}
\end{align}
where an arbitrary fixed point $x^*$ is incorporated in the coefficients $\underline{\theta}$ by algebraic transformation and comparison of coefficients.\\
The estimation of the model parameters $\theta_i$ and $\sigma$ is realised via a Markov Chain Monte Carlo (MCMC) method to reconstruct the full posterior probability density function (PDF) of the drift slope $\zeta$ and the noise level $\sigma$. The starting point is Bayes' theorem
\begin{equation}
p(\underline{\theta} |\underline{x}, \mathcal{I}) = \frac{p(\underline{x}|\underline{\theta}, \mathcal{I}) \cdot p(\mathcal{M}|\mathcal{I})}{p(\underline{\theta} | \mathcal{I})}
\end{equation}
with the posterior PDF $p(\underline{\theta} |\underline{x}, \mathcal{I})$, the likelihood $p(\underline{x}|\underline{\theta}, \mathcal{I})$, the prior $p(\mathcal{M}|\mathcal{I})$, and the evidence $p(\underline{\theta} | \mathcal{I})$ which accounts for the normalization of the posterior. The model parameters are denoted by $\underline{\theta}$, the time series data by $\underline{x}$, the background information by $\mathcal{I}$, and the model by $\mathcal{M}$.\\ 
The short-term propagator
\begin{align}
p(x, t=t' + \Delta t | x', t') = \frac{1}{\sqrt{2\pi g^2(x',t')\Delta t }}  \exp\left(-\frac{[x - x' - h(x',t')\Delta t]^2}{2 g^2(x',t')\Delta t}\right)
\end{align}
for subsequent times $t$ and $t'$ with $\tau = t - t' \longrightarrow 0$ can be derived from the Langevin equation if the difference $x - x'$ in the exponential expression is approximately defined by the first differences of a given time series. It represents the likelihood.\\
The priors are chosen to reflect the situation of no or just poor prior information. This guarantees the determination of the posterior mainly due to the available data instead of strong prior assumptions. More restrictive priors would be ill-advised in the subsequent analyses, since we have limited information about the model parameters describing the melt dynamics of the GrIS. For this reason, we assume an invariant prior~\cite{b:linden2014} for a straight line, i.e., for the intercept $\theta_0$ and the slope $\theta_1$. It is given by 
\begin{align}
p_{\rm prior}(\theta_0,\theta_1) = \frac{1}{2\pi (1+\theta_1^2)^\frac{3}{2}}
\end{align}
in a broad range of $[-50,50]$. Note that we denote $p_{\rm prior}(\theta_0,\theta_1)$ even if there is no explicit dependency on the intercept $\theta_0$. In the chosen variable representation, the slope $\theta_1$ is uniformly distributed in $\sin(\beta)$ with the angle $\beta$ between the x-axis and the straight line. It can be shown that this comes from the model assumption of an intercept $\theta_0$ (cf. Refs.~\cite{a:VanderPlas2014, a:Jaynes1999, b:linden2014}). For the noise level $\sigma$ in $[0,50]$ the invariant Jeffreys' scale prior~\cite{b:linden2014}
\begin{align}
p_{\rm prior}(\sigma ) = \frac{1}{\sigma}
\end{align}
is chosen because it is almost uninformative. Furthermore, it is taken care that the parameters of the higher orders are initially able to contribute in a similar magnitude to the deterministic dynamics as the linear ones by the broad Gaussian priors
\begin{align}\label{eq:gaussian prior}
&p_{\rm prior}(\theta_2) = \mathcal{N}(\mu = 0, \sigma_{\theta_2} = 4) \\ &p_ {\rm prior}(\theta_3) = \mathcal{N}(\mu = 0, \sigma_{\theta_3} = 8)
\end{align}
with mean $\mu$ and standard deviations $\sigma_{\theta_i}$. The MCMC affine-invariant ensemble sampler of the \textit{emcee}~\cite{a:Foreman-Mackey2013} python package is used to compute the posterior. Based on the estimated joint posterior PDF $p(\underline{\theta}|\underline{x},\mathcal{I})$, the parameters $\underline{\theta}$ are sampled and corresponding drift slopes $\zeta$ in the fixed point $x^*$ are calculated by marginalization:
\begin{linenomath*}
\begin{equation}
    p(\zeta |\underline{d},\mathcal{I}) = \int \!\! p(\underline{\theta},\sigma |\underline{d},\mathcal{I}) \delta\left( \zeta - \left.\frac{\text{d}h(x)}{\text{d}x}\right\vert_{x = x^*}\right) \text{d}\underline{\theta} \text{d}\sigma .
\end{equation}
\end{linenomath*}
\\
The credibility intervals (CIs) of the slopes and noise levels are defined as the $\SI{16}{\percent}$ to $\SI{84}{\percent}$- and $\SI{1}{\percent}$ to $\SI{99}{\percent}$-percentiles of the corresponding posterior PDFs. It is computed from a kernel density estimate of the corresponding PDFs. The kernel density estimation is performed with $scipy.stats.gaussian\_ kde$~\cite{a:Virtanen2020} using Silverman's rule of thumb to determine the kernel bandwidth.

\section{Results}
The BLE analyses are performed for two preprocessing scenarios: first, without the log transformation applied in the original article~\cite{a:Boers2021}, and second, with the log transformation, but using the same Gaussian kernel detrending (bandwidth $\sigma_{\rm k} = \SI{30}{a}$) to guarantee an easy comparison of the results (cf. Supplementary Information (SI)~\cite{SI:greenland}). The results of the analysis are presented in this section.\\
Principally, the log transformation addresses the skewed distribution of the raw data, on which the AR1 $\rho_{1}$ and STD $\hat{\sigma}_{\rm d}$ statistics are computed in the original article~\cite{a:Boers2021}. However, the BLE results for the CWG melt rates are almost equivalent across both preprocessing scenarios. Beyond that, the BLE method essentially involves calculating the first differences per window. The skewness of these differences is non-significant (cf. SI~\cite{SI:greenland}) for all windows in both datasets. Moreover, the skewness test results for the entire datasets of the logarithmic (~\makebox[0pt][l]{\raisebox{-0.4ex}{\includegraphics[width=0.6em]{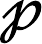}}}~~~~$=0.266$) and NU first differences (~\makebox[0pt][l]{\raisebox{-0.4ex}{\includegraphics[width=0.6em]{calp.pdf}}}~~~~$=0.192$) are non-significant at the $\SI{95}{\percent}$ confidence level. Against this background, if the underlying PDFs were skewed, the effect is likely to be negligible.\\
\subsection{Analysis of Melt Rate Dynamics}
\label{subsec: melt rate analysis}
\begin{figure}
\includegraphics[width=1.\textwidth]{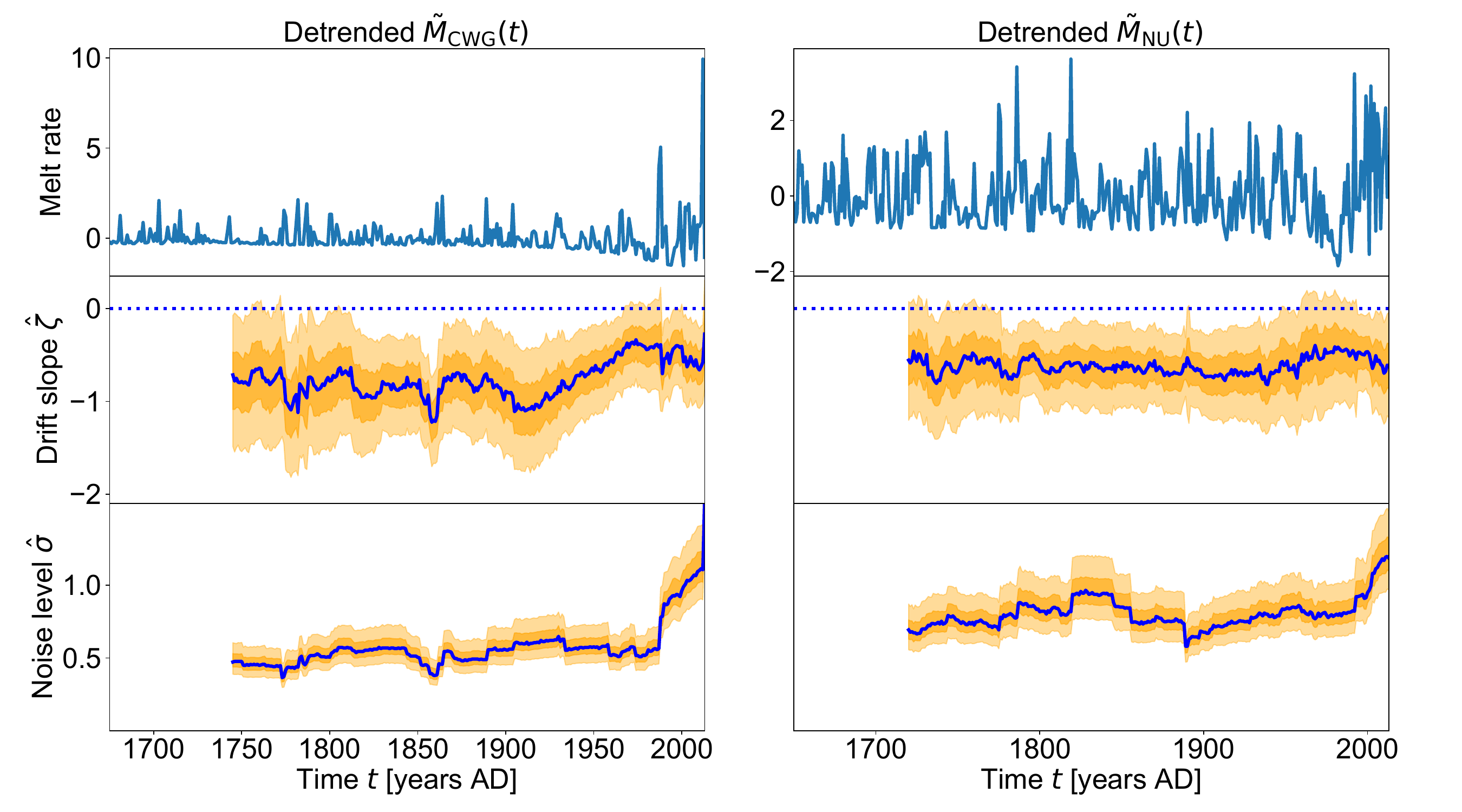}
\caption[Bayesian Langevin Model Estimation Results of the Greenland Icesheet Melt Rates]{BLE results for the GrIS melt rates extracted from stacked ice cores in CWG and from an ice core of the NU peninsula. The CWG drift slopes $\hat{\zeta}$ start to leave their stable plateau around 1914, which may be related to the ongoing industrialization in the late 19th century. By approximately 1975, a new, less resilient state is reached. Furthermore, the BLE resolves the increasing dominance of fast-scale perturbations beginning in 1987 (CWG) and 1992 (NU), implying a common driver of these dynamics. Large-scale fluctuations in atmospheric circulation~\cite{a:Slater2021} are a potential candidate for this driver. It is important to interpret these estimates with caution, as the model parameterisations certainly exhibit imperfections. Nevertheless, given the high degree of comparability regarding model assumptions and raw data aggregation, certain quantitative deductions are well justified. The drift slope and noise level estimates of the CWG melt rates in the higher regions of the main GrIS have approached the values of the NU peninsula, which is located at a lower elevation and in the coastal area, without a direct connection to the main GrIS. This might imply that the GrIS melt dynamics and conditions become more similar to those of the more variable coastal zones. However, further research is necessary to validate these hypotheses.}
\label{fig: greenland BLE}
\end{figure}
In Figure~\ref{fig: greenland BLE}, the BLE results for the CWG and NU melt rates $\tilde{M}_{\rm CWG}(t)$ and $\tilde{M}_{\rm NU}(t)$ are presented. The window size $N_{\rm w}=70$ data points is chosen analogously to Ref.~\cite{a:Boers2021}, and windows are shifted by one year. The BLE drift slopes $\hat{\zeta}$ clearly indicate a rather homogeneous state of higher resilience before 1914. From that point onward, the drift slopes increase until roughly 1975, suggesting that the GrIS reaches a plateau of weaker resilience around this time. The starting point of the gradual loss of stability, considering some time lag due to the rolling window approach, is in good agreement with previous findings of increased GrIS melt rates during the course of ongoing industrialization in the late 1800s.~\cite{a:Trusel2018} Moreover, the less resilient plateau observed after 1975 is associated with the stabilized AR1 $\hat{\rho}_{1}$ already observed by Boers and Rypdal~\cite{a:Boers2021}. Interestingly, the BLE results suggest that the recently less resilient state of CWG melt rates is now approaching the nearly constant resilience level of the NU melt rates.\\
\begin{figure}[h!]
\includegraphics[width=\textwidth]{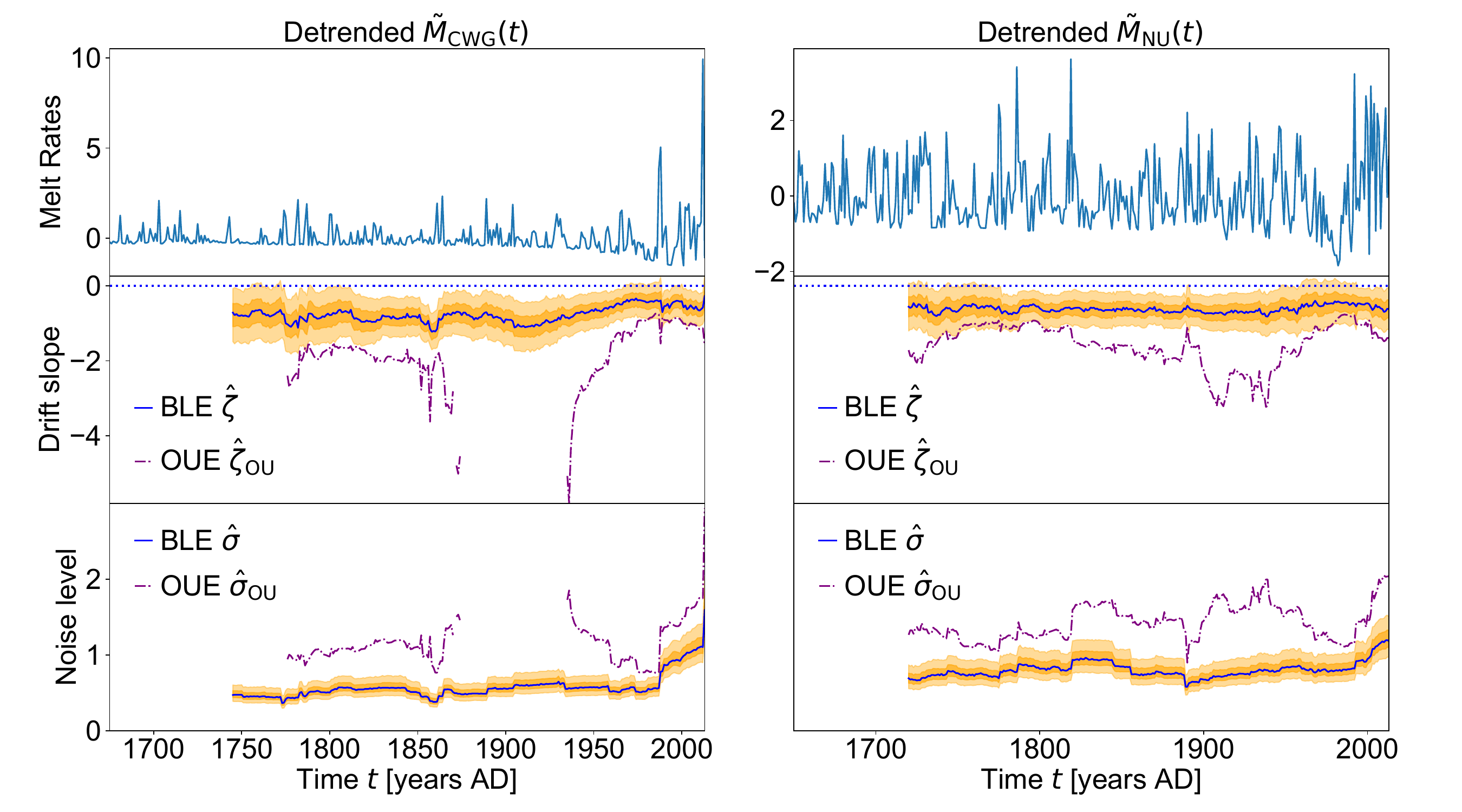}
\caption[The Greenland Ice Sheet and the Ornstein-Uhlenbeck Approximation]{Comparison of the OUE and the BLE for the melt rates of the GrIS extracted from stacked ice cores in CWG and an ice core of NU. The BLE resolves resilience changes of the CWG melt rates starting in the 1910s and highlights the increasing influence of fast-scale processes from the 1970s onward. In contrast, the CWG OUE exhibit significant variability and are undefined for periods with AR1 $\rho_{1}<0$. Similarly, the NU OUE results fluctuate heavily. Since it is not straightforward to define CIs for these OU estimates, their interpretation is strongly limited. The comparison of the log-transformed versions of the data yields almost perfectly identical results (cf. SI~\cite{SI:greenland}).}
\label{fig: greenland OU}
\end{figure}
Such a rough quantitative comparison of the drift slopes across both time series is well justified for several reasons. On the one hand, we consider the \textit{same} quantity, i.e., melt rates, with exactly the \textit{same} time resolution from the NU peninsula and CWG. On the other hand, we apply the \textit{same} parametric model, window sizes, and shifts to both datasets.\\
CWG is part of the inland main GrIS, with CWG core locations at altitudes of $\sim \SIrange{2300}{2525}{\meter}$, i.e., within the GrIS accumulation zone ($\gtrsim\SI{1200}{\meter}$)~\cite{a:Trusel2018}. In contrast, the NU core is taken from an ice cap in the coastal region without direct connection to the main GrIS, lies at a lower elevation ($\sim\SI{1700}{\meter}$)~\cite{a:Das2018}, and exhibits a greater overall magnitude of melt~\cite{a:Trusel2018}. Additionally, Trusel et al.~\cite{a:Trusel2018} found that the CWG melt stack and NU melt rates have high predictive power for large parts of the GrIS.\\ 
Considering these facts, the BLE results suggest that larger parts of the GrIS approach the less stable level of the NU peninsula ice cap, and their melt dynamics might even become more similar to those of the NU ice cap. This reasoning is plausible in light of the generally higher melt rates in the coastal area, including the NU peninsula, and the recent net mass loss of the GrIS. Thus, the melt dynamics of the NU ice cap, which depend more strongly on air temperature variability~\cite{a:Trusel2018}, may serve as a prototype for the future developments over the entire GrIS up to a certain extent.\\
Perhaps the most intriguing BLE result is the strikingly similar evolution and magnitude of the noise levels $\hat{\sigma}$ for both time series. If we compare the agreement of the CWG and NU noise levels during the stable period---i.e., before 1987 for CWG and before 1992 in the NU peninsula---to the available average from the 2010s, we see that this agreement increased from approximately $\SI{67}{\percent}$ to $\SI{97}{\percent}$. This further supports the notion of converging dynamics not only on slow deterministic but also on fast stochastic scales. Furthermore, the overall trend similarities imply that the melt dynamics of CWG and NU melt rates likely share common fast-scale stochastic driving processes. These processes may include large-scale fluctuations in atmospheric circulation, which have become more dominant in the recent decades, as noted by Slater et al.~\cite{a:Slater2021}. The authors additionally reconstructed runoffs and found an approximately $\SI{60}{\percent}$ increase in runoff variability in the 2010s compared to the previous three decades. A comparison of the average noise levels from available 2010s data similarly shows an increase about $\SI{50}{\percent}$ for CWG and $\SI{30}{\percent}$ for NU melt rates. Since melt rates contribute to the overall runoff dynamics, the roughly matching magnitudes of the increase in variability provide further evidence supporting the hypothesis that large-scale fluctuations in atmospheric circulation are partly responsible for the recent parallel  rise in noise levels in CWG and NU melt rates.\\ 
\subsection{Limits of the Ornstein-Uhlenbeck Approximation}
\label{subsec: greenland BLE OUE}
The GrIS melt rates $\tilde{M}_{\rm CWG}(t)$ and $\tilde{M}_{\rm NU}(t)$ serve as a prototypical real-world example to demonstrate the superiority of the developed BLE over common non-parametric and parametric leading indicators. As outlined in the previous section, the BLE provides more detailed insights into the melt rate dynamics from coastal to inland areas of the GrIS compared to the AR1 and STD statistics used by Boers and Rypdal~\cite{a:Boers2021}. The OUE metrics, derived from AR1 and STD by the first approximation of an OU process, are shown to be insufficient for capturing the BLE results. A comparison is provided in Figure~\ref{fig: greenland OU}.\\
The BLE drift slopes $\hat{\zeta}$ from the CWG dataset suggest an nearly unaltered state until around 1914, when the melt rate dynamics shift to a less stable new plateau, reached by approximately 1975. In contrast, the CWG OUE drift slopes $\hat{\zeta}_{\rm OUE}$ exhibit strong fluctuations without well-defined plateaus or trends. The positive trend from $-\infty$ to roughly $-0.85$ starting around 1936 is challenging to interpret without simultaneously considering the BLE, as the preceding intervals are either undefined or highly variable. The undefined estimates for certain time stamps $t$ stem from the multidecadal variability~\cite{a:Ghil1991} of AR1 $\rho_{1}$, which is a known feature of the CWG melt rates: In periods with AR1 $\rho_{1}<0$, the OUE breaks down, and the estimates diverge (cf. Equations~\ref{eq:OU indicator} and Figure~\ref{fig: greenland OU}). Similarly, the CWG OUE noise levels $\hat{\sigma}$ do not resolve the nearly stable plateau before the increased dominance of fast-scale processes starting around 1987. The undefined estimates again correspond to the periods with $\rho_{1}<0$. Likewise, the NU OUE estimates fluctuate heavily over the years, making it almost impossible to infer reliable trends. Log-transforming the data leaves these findings unaltered (cf. SI~\cite{SI:greenland}).
\FloatBarrier
\section{Conclusion}
The BLE analysis of GrIS melt rates underscores the method's widespread potential. The method reveals a trend of diminished resilience for the GrIS in line with other studies~\cite{a:Boers2021,a:Trusel2018}. Additionally, the BLE estimates from the inland GrIS data are approaching the values observed in coastal areas. This suggests the possibility that the melt dynamics in higher regions of the main GrIS may converge toward those of the more variable lower coastal areas. Furthermore, there is a very strong correspondence between the BLE noise estimates from the coastal NU peninsula and the main GrIS CWG melt rates, with a strongly increasing impact of these dynamics emerging around the 1990s. This may indicate a common fast-scale driver of intrinsic melt rate stochasticity, with large-scale fluctuations of atmospheric circulation being a potential candidate~\cite{a:Slater2021}. These relationships are revealed through the robust BLE and cannot be reproduced employing AR1, STD, or the inadequate approximation of the OUE.

\subsection*{Data and Code Availability Statement}
The analysed melt rate time series from the GrIS are freely available at \url{https://www.nature.com/articles/s41586-018-0752-4#Sec14}. The conducted BLE analyses can be reproduced using the open-source Python package \textit{antiCPy} hosted at \url{https://github.com/MartinHessler/antiCPy} under a \textit{GNU General Public License v3.0} complemented by a detailed documentation at \url{https://anticpy.readthedocs.io}.

\subsection*{Author Contributions}
MH and OK designed research. MH implemented the method, preprocessed data, conducted data analysis and research, and designed graphics. MH wrote the manuscript in consultation with OK.

\subsection*{Additional Information}
This research received no external funding. The authors declare no competing interests.

\subsection*{Acknowledgments}
M.H. thanks the Studienstiftung des deutschen Volkes for a scholarship including financial support.

\bibliographystyle{unsrtnat}
\bibliography{references.bib}

\end{document}